\documentclass[aps,prbr,reprint,superscriptaddress,groupedaddress,citeautoscript,floatfix]{revtex4-1}
\usepackage{amsmath,amssymb,amsfonts,bbm,textcomp}
\usepackage[english]{babel}
\usepackage[utf8]{inputenc}
\usepackage[T1]{fontenc}
\usepackage{times}
\usepackage[shortcuts]{extdash}
\usepackage{graphicx}
\usepackage{hyperref}
\hypersetup{pdftitle=Controlling Majorana states in topologically inhomogeneous superconductors, pdfauthor=Pasquale Marra}
\usepackage[capitalize]{cleveref}
\newcommand{\pf}{\operatorname{pf}}
\newcommand{\sgn}{\operatorname{sgn}}
\newcommand{\up}{{\uparrow}}
\newcommand{\down}{{\downarrow}}
\newcommand{\nodag}{{\phantom{\dag}}}

\newcommand{\meanbsq}{{\langle b^2\rangle}}
\newcommand{\meanb}{{\langle b\rangle}}
\newcommand{\citesupp}{Supplemental Material}

\begin{document}
\title{Controlling Majorana states in topologically inhomogeneous superconductors}

\author{Pasquale Marra}
\email{pasquale.marra@spin.cnr.it}
\author{Mario Cuoco}
\affiliation{CNR-SPIN, I-84084 Fisciano (Salerno), Italy}
\affiliation{Department of Physics ``E.\,R.\, Caianiello'', University of Salerno, I-84084 Fisciano (Salerno), Italy}


\begin{abstract}
Majorana bound states have been recently observed at the boundaries of one-dimensional topological superconductors.
Yet, controlling the localization of the Majorana states, which is essential to the realization of any topological quantum device, is an ongoing challenge.
To this end, we introduce a mechanism which can break a topologically homogeneous state via the formation of topological domains, and which can be exploited to control the position of Majorana states.
We found in fact that in the presence of amplitude-modulated fields, contiguous magnetic domains can become topologically inequivalent and, as a consequence, Majorana states can be pinned to the domain walls of the magnetic structure.
The formation of topological domains and the position of Majorana states can be externally controlled by tuning an applied field (e.g., magnetic or gate).
\end{abstract}

\maketitle

\paragraph*{Introduction}

The experimental observation of Majorana bound states (MBS)~\cite{Kitaev2001} in topological superconductors~\cite{Hasan2010,Zhang2011,Alicea2012,Beenakker2013} marks the first milestone on the pathway toward topological quantum computation~\cite{Kitaev20032}.
In particular, topological superconductivity has been observed in proximized nanowires with strong spin-orbit coupling (SOC)~\cite{Mourik2012,Das2012,Rokhinson2012,Deng2012,Finck2013,Churchill2013}, and in ferromagnetic atomic chains on the surface of a superconductor~\cite{Nadj-Perge2014,Pawlak2016,Ruby2015}.
In general, conventional superconductivity can be turned topological by the presence of a uniform magnetic field and intrinsic SOC~\cite{Lutchyn2010,Oreg2010}, an intrinsic ferromagnetic order and SOC~\cite{Brydon2015}, or by a noncollinear spatially-dependent magnetic field~\cite{Choy2011,Klinovaja2012PRL,Nadj-Perge2013,Chen2015}.
In particular, a very promising system is represented by chains of Yu-Shiba-Rusinov states~\cite{Yu1965,Shiba1968,Rusinov1969} induced by magnetic atoms on the surface of conventional superconductors with ferromagnetic~\cite{Brydon2015}, antiferromagnetic~\cite{Heimes2014}, or helimagnetic textures~\cite{Klinovaja2013,Braunecker2013,Vazifeh2013,Pientka2013,Nakosai2013,Pientka2014,Kim2014,Rontynen2014,Poyhonen2014,Chen2015,Schecter2016}.

Nevertheless, the implementation of a reliable braiding scheme~\cite{Alicea2011,Clarke2011,Halperin2012}, which is essential to the realization of topological quantum devices, is an ongoing challenge.
A necessary prerequisite is the ability to control the position of MBS, which are localized at the boundaries between topologically trivial and nontrivial domains.
In the proposed schemes~\cite{Lutchyn2010,Oreg2010,Choy2011,Klinovaja2012PRL,Nadj-Perge2013}, these boundaries are determined by the experimental setup and geometry, and can be manipulated via a gradient of the gate voltage~\cite{Alicea2011} or magnetic field~\cite{Kim2015}, via gate-tunable valves~\cite{Aasen2016}, via the magnetic flux in Josephson junctions~\cite{Heck2012}, via external magnetic fields in the presence of an helimagnetic order and SOC~\cite{Li2016}, or by controlling the magnetic texture in two-dimensional electron gases~\cite{Fatin2016}.

In this Rapid Communication, we introduce a physical mechanism which can be exploited to locally break a homogeneous superconducting state into inhomogeneous \emph{topological} domains and to control the position of MBS\@.
We found in fact that, in the presence of amplitude-modulated magnetic fields, the superconducting gap and the topological invariant strongly depends on the phase-offset $\varphi$ of the periodic texture.
Consequently, contiguous magnetic domains can become topologically inequivalent and MBS can be pinned to the domain walls of the magnetic structure.
This leads to the emergence of a topologically inhomogeneous state characterized by contiguous inequivalent domains, in an otherwise homogeneous superconducting state.
The main advantage is that the formation of topological domains, and thus the position of MBS, can be externally controlled by tuning an applied \emph{uniform} field.
Differently from other proposals~\cite{Alicea2011,Clarke2011,Halperin2012,Kim2015}, this mechanism does not rely on the manipulation of gradients of the field intensity.

\paragraph*{The model}

\begin{figure}
\centering
\includegraphics[scale=1,resolution=600]{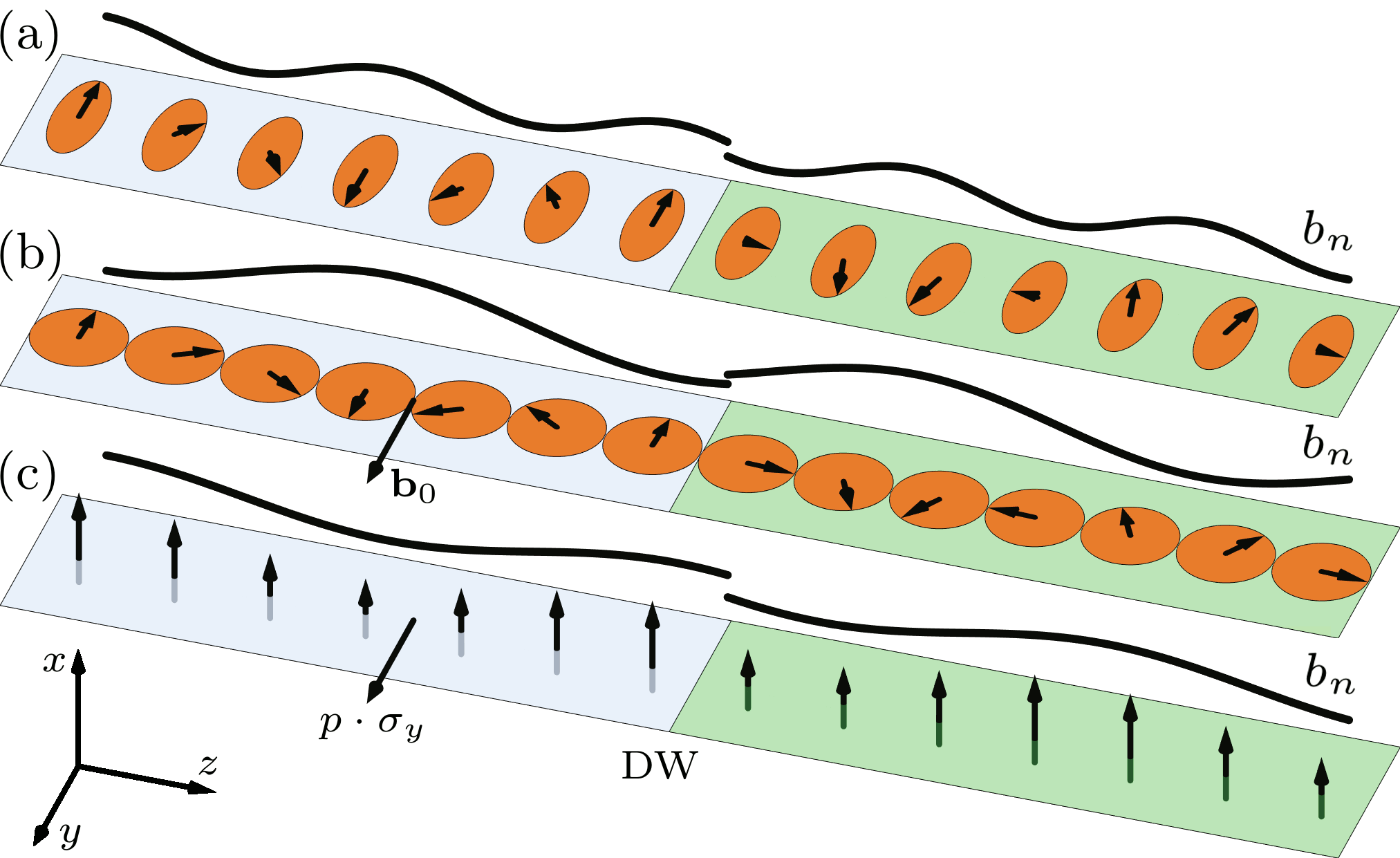}\\[-1mm]
	\caption{%
Topologically inhomogeneous superconductors can be realized in superconducting systems with amplitude-modulated magnetic fields, e.g., (a) elliptical helical field, (b) circular helical field superimposed with an externally applied field $\mathbf{b}_0$, or (c) collinear amplitude-modulated field with intrinsic SOC\@.
The field amplitude $b_n$ is not constant, but periodically modulated along the chain.
These systems can exhibit domain walls (DW) which break the spatial periodicity of the field.
}
	\label{fig:System}
\end{figure}

Noninteracting $s$-wave superconductors with periodically amplitude-modulated magnetic (Zeeman) fields can be described by a Bogoliubov-de~Gennes (BdG) tight-binding Hamiltonian in the form
\begin{gather}
\mathcal{H}=
\frac12\sum_n
\boldsymbol\Psi_n^\dag
\!\cdot\!
\begin{bmatrix}
2t-\mu+\mathbf{b}_n\!\cdot\!\boldsymbol{\sigma} & \imath\sigma_y\Delta \\
(\imath\sigma_y\Delta)^\dagger & -(2t-\mu+\mathbf{b}_n\!\cdot\!\boldsymbol{\sigma}^*)
\end{bmatrix}
\!\cdot\!
\boldsymbol\Psi_n
\nonumber\\
-
\frac12\sum_n
\boldsymbol\Psi_n^\dag
\!\cdot\!
\begin{bmatrix}
t\Omega&0\\
0&-t\Omega
\end{bmatrix}
\!\cdot\!
\boldsymbol\Psi_{n+1}
+\text{H.c.},
\label{eq:Hamiltonian}
\end{gather}
where $\boldsymbol\Psi_n^\dag=[c^\dag_{n\up},c^\dag_{n\down},c^\nodag_{n\up},c^\nodag_{n\down}]$ is the Nambu spinor with $c^\nodag_{n\up\down}$ and $c^\dag_{n\up\down}$ the electron annihilation and creation operators, $\boldsymbol{\sigma}=[\sigma_x,\sigma_y,\sigma_z]$ the vector of Pauli matrices, $\mu$ the chemical potential, $t$ the hopping parameter, $\Omega=\mathbbm{1}+\imath\lambda \sigma_y$ with $\lambda$ the intrinsic SOC, $\Delta$ the superconducting gap, and $\mathbf{b}_n$ the magnetic field on the site $r_n$.
We assume for the sake of simplicity that the field modulation is commensurate with the lattice, i.e., it has spatial frequency $\theta=2\pi p/q$ with $p,q\in\mathbb{Z}$ coprimes.
Topologically nontrivial gapped phases are realized in the presence of either a collinear magnetic field (constant direction) with intrinsic SOC $\lambda\neq0$~\cite{Lutchyn2010,Oreg2010} or of a noncollinear field (nonconstant direction) with $\lambda=0$~\cite{Choy2011,Nadj-Perge2013,Chen2015}, or both~\cite{Klinovaja2012PRL}.
In the second case in fact, the variation of the field direction is unitarily equivalent to an effective SOC~\cite{Braunecker2010}.
In order to see this, one can rotate the $z$ axis of the spin basis at each lattice site to the field direction via a unitary transformation~\cite{Choy2011,Kjaergaard2012} $U_n$.
If the SOC vanishes in fact, the transformed BdG Hamiltonian~\cite{Choy2011} coincides with Hamiltonian~\eqref{eq:Hamiltonian} after the substitutions
\begin{equation}
\Omega\to \Omega_n=U_{n}^{\dagger}U_{n+1},\qquad\mathbf{b}_n\cdot\boldsymbol\sigma\to b_n\sigma_z.
\label{eq:HamiltonianSubstitution}
\end{equation}
This mandates the presence of an effective collinear Zeeman field along the $z$ axis with the same amplitude as the original field, a renormalized kinetic term (diagonal terms of $\Omega_n$), and an effective SOC (off-diagonal terms of $\Omega_n$) which is nonzero if the original field $\mathbf{b}_n$ is noncollinear~\cite{Choy2011}.
In the presence of a finite magnetic field and SOC (effective or intrinsic), Hamiltonian~\eqref{eq:Hamiltonian} exhibits gapped phases which break time-reversal and chiral symmetries.
These are the necessary ingredients which allow the realization of $s$-wave topological superconductors~\cite{Lutchyn2010,Oreg2010,Choy2011,Nadj-Perge2013} characterized by a nontrivial $\mathbb{Z}_2$ topological invariant.
This invariant coincides with the fermion parity~\cite{Kitaev2001,Tewari2012,Budich2013} $\mathcal{P}=\prod_{k=-k}\sgn[F(k)]$, defined in terms of the Pfaffians $F(k)=\pf[\imath\tau_x H(k)]$ at the time-reversal invariant momenta $k=-k$, being $H(k)$ the BdG Hamiltonian in momentum space and $\tau_x$ the first Pauli matrix in particle-hole space.
In the low-energy sector, this system is equivalent to a $p$-wave topological superconductor~\cite{Kitaev2001,Lutchyn2010,Oreg2010,Choy2011}.

\paragraph*{Dependence on the phase-offset}

In the infinite-chain limit (i.e., neglecting finite-size effects) and for vanishing SOC ($\lambda=0$), the Hamiltonian~\eqref{eq:Hamiltonian} is invariant under \emph{global} rotations of the spin basis or, equivalently, of the Zeeman field.
In order to see this, consider the case of a helical field with direction uniformly varying within the plane $yz$, i.e., $\mathbf{b}_n=b [\sin{(n\theta+\varphi)}\hat{\mathbf{y}} + \cos{(n\theta+\varphi)}\hat{\mathbf{z}} ]$ where $\theta$ is the angular variation of the field direction between adjacent sites, and $\varphi$ the phase-offset describing the boundary-offset of the magnetic texture.
A global rotation of the spin basis corresponds to a variation of the phase-offset.
Moreover, in this case the transformed Hamiltonian obtained via \cref{eq:HamiltonianSubstitution} does not depend explicitly on the phase-offset $\varphi$, since both the effective Zeeman and SOC terms become uniform, i.e., $b_n \hat{\mathbf{z}}=b \hat{\mathbf{z}}$ and $\Omega_n=U_n^\dagger U_{n+1}=\cos{(\theta/2)}+\imath\sigma_x\sin{(\theta/2)}$ (see \citesupp).
The phase-offset $\varphi$ is thus immaterial, since it can be absorbed by a unitary transformation.
Hence, the system exhibits a global $U(1)$ gauge-invariance with respect to the phase-offset $\varphi$ as a consequence of the more general $SO(3)$ global spin-rotational symmetry.
Thus, in the case of circular helical fields without SOC, the bulk electronic spectrum does not depend on the boundary-offset of the magnetic texture (except for energy contributions at the edges of the chain).

Nevertheless, if the SOC vanishes, an amplitude-modulated field can break the phase-offset invariance even without breaking the full spin-rotational symmetry.
Consider, e.g., a noncollinear amplitude\-/modulated field realized via an elliptical helical field $\mathbf{b}_n=b_y\sin{(n\theta+\varphi)}\hat{\mathbf{y}} + b_z\cos{(n\theta+\varphi)}\hat{\mathbf{z}}$, with $b_y\neq b_z$, shown in \cref{fig:System}(a), or via circular helical field superimposed with a coplanar uniform field $\mathbf{b}_n=b_{yz} \left[\sin{(n\theta+\varphi)}\hat{\mathbf{y}} + \cos{(n\theta+\varphi)}\hat{\mathbf{z}}\right]+\mathbf{b}_0$ with $\mathbf{b}_0$ in the $yz$ plane, shown in \cref{fig:System}(b).
As one can verify, in these cases the field amplitude is not uniform, but periodically modulated along the chain as $b_n=\sqrt{\meanbsq+{\delta b}^2 \cos{[j(n\theta+\varphi)]}}$ with $j=2$ and 1, respectively, in the case of elliptical and circular helical field.
Hence the boundary phase-offset $\varphi$ cannot be absorbed by unitary rotations, since the phase-offset affects not only the direction, but also the amplitude\-/modulation of the field.
Analogously, one can consider a collinear and amplitude\-/modulated magnetic field, e.g., $\mathbf{b}_n=[\meanb+\delta b\cos{(n\theta+\varphi)}]\hat{\mathbf{z}}$, shown in \cref{fig:System}(c), with a finite SOC $\Omega=\mathbbm{1}+\lambda\imath\sigma_y$.
Notice that on a discrete lattice, the field texture is not invariant under translations $n\to n-\varphi/\theta$ unless $\varphi/\theta$ is an integer.
Hence, the phase-offset cannot be absorbed by spatial translations in the general case.

In all these systems, the boundary phase-offset cannot be gauged away by any unitary transformation, if the amplitude-modulation ${\delta b}$ is finite.
Thus, the energy spectrum will depend explicitly on the offset of the magnetic texture at the boundary.
This dependence is not due to a change of the average field per unit cell, which in fact does not depend on the phase-offset $\varphi$ in the cases considered here (harmonic modulation), as shown in the \citesupp.
Nevertheless, changes of the phase-offset $\Delta\varphi=m\theta$ which are integer multiples of the angle $\theta$, are equivalent to a discrete lattice translation $n\rightarrow n-\Delta\varphi/\theta$.
This mandates a periodicity of the Hamiltonian (up to lattice translations) in the phase-offset $\varphi$ with period $\Delta\varphi=2\pi/q$ (see also Ref.~\onlinecite{Marra2015}).
Moreover, the system is also periodic in the momentum with the same period (see \citesupp), which results in a reduced Brillouin zone given by $[0,2\pi/q]$.
The global $U(1)$ continuous symmetry (arbitrary variations $\Delta\varphi$ of the phase-offset) is broken down into a discrete symmetry (periodicity in the phase-offset $\Delta\varphi=m\theta$ integer).
This is analogous to the case of a crystal lattice which breaks the continuous translational symmetry of free space.
Notice that the boundary-dependence can be obtained not only via magnetic fields, but more generally via any amplitude-modulated field as, e.g., electric field~\cite{Gangadharaiah2012,Marra2015,Park2016} (charge-density waves) or strain-induced SOC fields~\cite{Wang2004,Bernevig2005,Gentile2013,Calleja2015,Klinovaja2015}.

\begin{figure}
\centering
\includegraphics[scale=1,resolution=600]{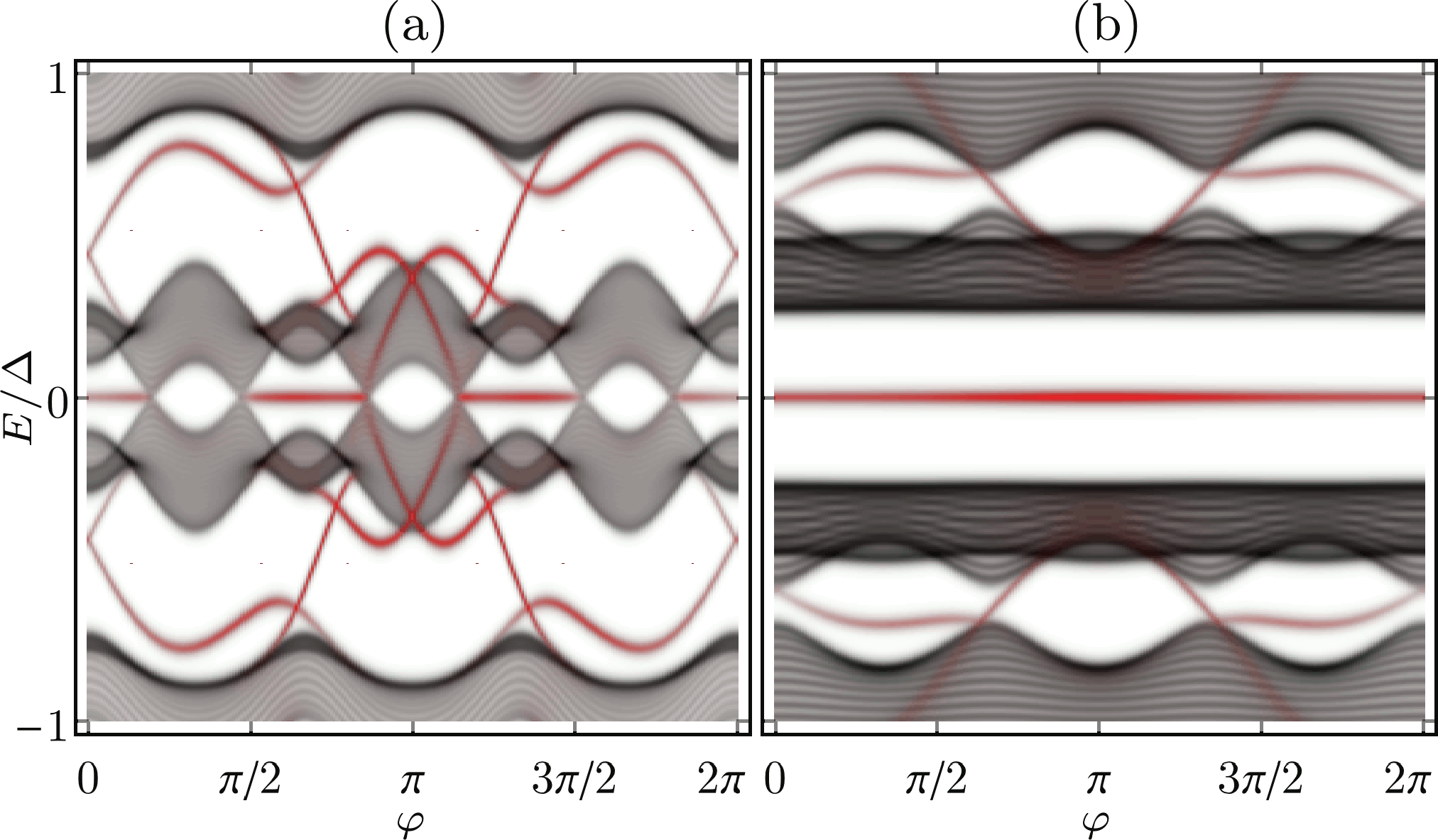}\\[-1mm]
	\caption{%
DOS in the bulk (dark) and at the edges (color) of the system described by Hamiltonian~\eqref{eq:Hamiltonian} in the low-energy range as a function of the boundary phase-offset $\varphi$ in the case of a circular helical field with spatial frequency $\theta=2\pi/3$ and amplitude $b_{yz}$ superimposed with an applied field $\mathbf{b}_0$ [as in \cref{fig:System}(b)].
(a) Nontrivial gaps with MBS alternating with trivial gaps as a function of the phase-offset $\varphi$ obtained for $b_0/b_{yz}=0.75$ (${\delta b}^2/\meanbsq \approx1$).
(b) Nontrivial gap with MBS in the whole range $\varphi\in[0,2\pi]$ for $b_0/b_{yz}=0.3$ (${{\delta b}^2/\meanbsq} \approx0.5$).
Notice that in both cases the bulk energy levels are periodic in the phase-offset with period $\Delta\varphi=\theta=2\pi/3$.
We assume $\Delta=t/2$, $\mu=3t$, $b_{yz}=1.5t$, and $\mathbf{b}_0\parallel \hat{\mathbf{z}}$.
}
	\label{fig:Edges}
\end{figure}

\Cref{fig:Edges} shows the dependence on the boundary phase-offset $\varphi$ of the density of states (DOS) at low-energy spectra of a one-dimensional superconductor with an amplitude-modulated magnetic field, calculated from Hamiltonian~\eqref{eq:Hamiltonian}.
In particular, we considered a circular helical field with spatial frequency $\theta=2\pi/3$ superimposed with an applied uniform field [see \cref{fig:System}(b)], for two choices of the applied field.
The bulk energy spectrum depends periodically on the phase-offset $\varphi$ with period $\Delta\varphi=\theta$.

\paragraph*{Closing the particle-hole gap}

As we have shown, the energy spectrum of a superconductor in the presence of amplitude-modulated magnetic fields can depend explicitly on the offset $\varphi$ of the magnetic texture.
In fact, if the energy variation $\Delta E$ of the lowest-energy level is comparable with the superconducting pairing $\Delta$, the particle-hole gap may close and reopen at specific values of the boundary-offset $\varphi=\varphi^*$.
This is indeed the case shown in \cref{fig:Edges}(a), where the bulk particle-hole gap closes for certain values of the phase-offset $\varphi$.
Hence, the bulk properties can be deeply modified by the phase-boundary dependence:
A conventional nodeless $s$-wave superconductor becomes an unconventional nodal superconductor for specific values of the phase-offset.
In general, if the lowest-energy level closes and reopens the gap linearly in the phase-offset ($E\propto\varphi-\varphi^*$) at any of the two time-reversal invariant momenta $k=0,\pi/q$ in the reduced Brillouin zone $[0,2\pi/q]$, and if no additional degeneracies are present, the Pfaffian $F(k)$ will change its sign and the topological invariant shall change at $\varphi=\varphi^*$ (see also Ref.~\onlinecite{Marra2016}).
Fermion parity transitions do indeed occur if the amplitude-modulation $\delta b$ is large compared with the average magnetic field.
This can be verified by expanding the Pfaffian as a Fourier sum in the phase-offset.
For collinear amplitude-modulated fields $b_n=\meanb+\delta b\cos{(n\theta+\varphi)}$ [\cref{fig:System}(c)] with finite SOC, one obtains (neglecting higher harmonics)
\begin{equation}
F(k)\approx F_0(k)+C{\delta b}^{q}\cos{(q\varphi)}\quad \text{for}\quad k=0, \pi/q
\label{eq:Pfaffian}
\end{equation}
where $F_0(k)=\pm\sqrt{\prod_{m=0}^{q-1} \det{\left[ h(k+m\theta)\right]}}$, being $h(k)$ the Hamiltonian in momentum space with constant magnetic field ($\delta b=0$), and where $C$ is a constant prefactor (see \citesupp).
Similar equations can be obtained in the case of noncollinear amplitude-modulated fields.
Thus, if the particle-hole gap remains open for any value of the phase-offset $\varphi$, the fermion parity $\mathcal{P}\!=\!\sgn[F(0)F(\pi/q)]$ will change its sign at $\varphi\equiv\varphi_m^*\!=\!\{\arccos{[({\delta b}_c/{\delta b})^q]}\!+\!2\pi m\}/q$ for $m\in\mathbb{Z}$ if the amplitude-modulation is larger than a critical value ${\delta b}>{\delta b}_c\equiv\sqrt[q]{\min[|F_0(0)|,|F_0(\pi/q)|]/|C|}$.

Hence, for commensurate spatial frequencies $\theta=2\pi p/q$ ($p,q$ coprimes) with $q$ odd, a change of the phase-offset can drive a transition between trivial and nontrivial states if the amplitude-modulation ${\delta b}$ is larger than a critical value ${\delta b}_c$.
A remarkable consequence is that MBS can be created or annihilated by changing the phase-offset at the boundary.
This is indeed the key result of this Rapid Communication and the main property of topological superconductors in the presence of amplitude-modulated fields:
The particle-hole gap can be closed and reopened with a concurrent transition between trivial and nontrivial states by tuning only the phase-offset $\varphi$ of the magnetic texture, i.e., without changing the field average intensity.
In particular, \cref{fig:Edges}(a) shows the case where ${\delta b}>{\delta b}_c$, with nontrivial gaps ($\mathcal{P}=-1$) with MBS alternating to trivial gaps ($\mathcal{P}=1$) for different values of the phase-offset $\varphi\in[0,2\pi]$.
Alternatively, a nontrivial gap can span the whole interval $[0,2\pi]$, as shown in \cref{fig:Edges}(b) in the case that ${\delta b}<{\delta b}_c$, where MBS are present for any possible boundary configuration ($\mathcal{P}=-1$ for any choice of the phase-offset $\varphi$).

\paragraph*{Pinning of MBS}

The dependence of the topological properties on the offset of the magnetic texture can be exploited to control the localization of MBS in the presence of a domain wall which breaks the periodic rotation of the field direction, without changing its average intensity.
Consider in fact the case shown in \cref{fig:System}, where a domain wall at the lattice site $r_m$ separates two contiguous and homogeneous magnetic domains (left and right domains), which are characterized by a different value of the phase-offset, i.e., $\varphi_L$ and $\varphi_R$ with $\Delta\varphi_m\equiv\varphi_R-\varphi_L\neq0$.
If the field amplitude is constant and the SOC vanishes ($\lambda=0$), the spin-rotational symmetry of Hamiltonian~\eqref{eq:Hamiltonian} is unbroken. 
In this case, neglecting boundary/termination conditions and additional symmetry-breaking effects, the phase-offset can be gauged away by unitary rotations. 
Thus, under these assumptions, the two domains are unitarily and topologically equivalent, i.e., $\mathcal{P}(\varphi_L)=\mathcal{P}(\varphi_R)$, being either both trivial or nontrivial.
On the contrary, if the field is amplitude-modulated, the two domains are not unitarily equivalent anymore, since the phase-offset cannot be gauged away by unitary transformations.
Moreover, if the amplitude-modulation is large enough ($\delta b>{\delta b}_c$), the topological invariant will depend on the phase-offset:
In this case their fermion parity may differ, i.e., $\mathcal{P}(\varphi_L)\neq\mathcal{P}(\varphi_R)$ for certain values of the phase-difference $\Delta\varphi_m$.
Hence, two distinct and contiguous \emph{topological domains} emerge, separated by the domain wall of the underlying magnetic texture.
Consequently, MBS localize at the boundaries of the topologically nontrivial domain, i.e., at one edge of the chain and at the domain wall.
Notice that, for the sake of simplicity, the domain wall thickness is assumed to be equal to one interatomic distance. 
However, domain walls in magnetically ordered materials typically extend over several lattice sites: In this case the pinned MBS would be localized over a similar length scale.

\begin{figure}
\centering
\includegraphics[scale=1,resolution=600]{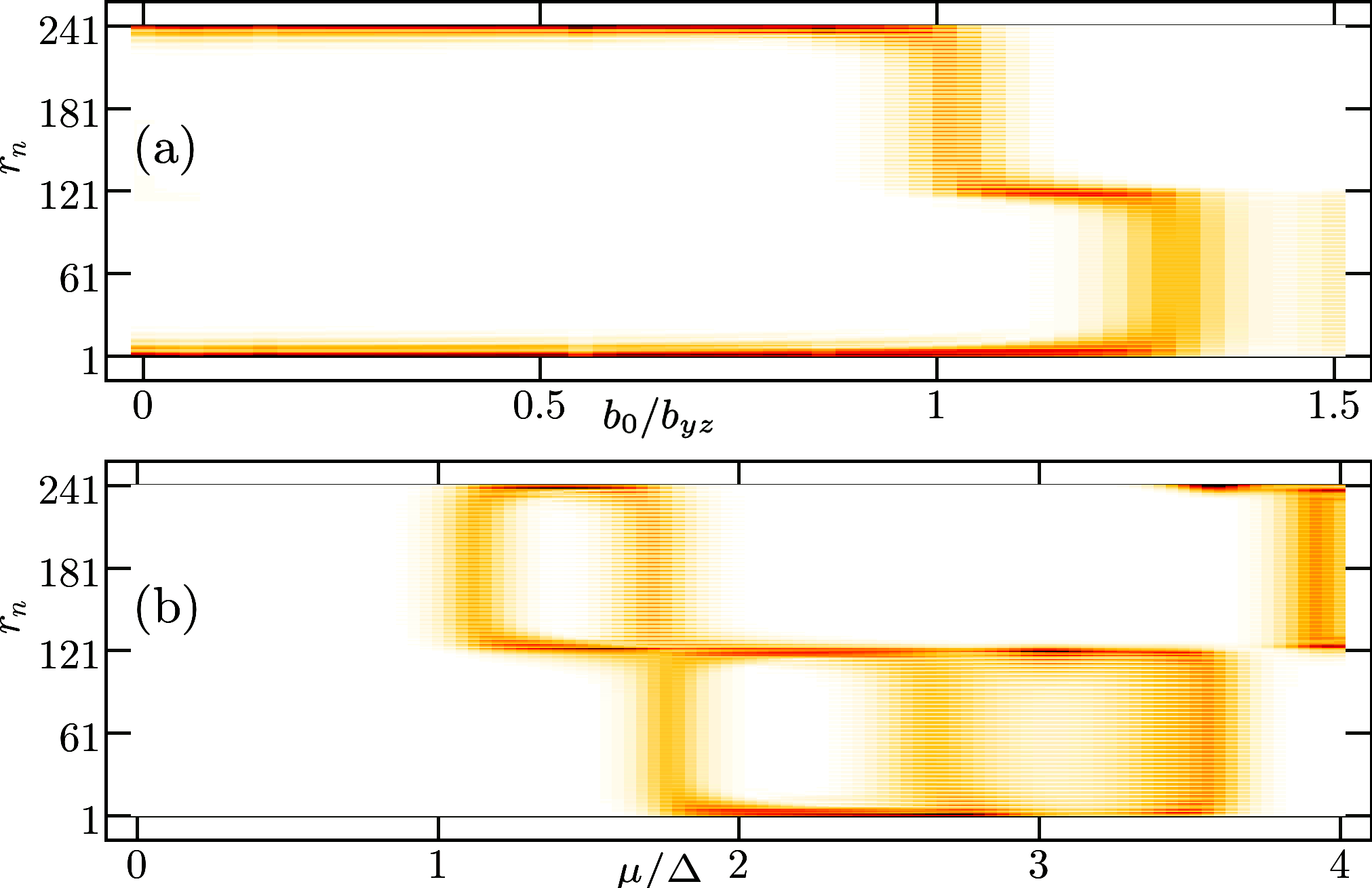}\\[-1.5mm]
	\caption{%
LDOS at zero energy in the presence of a domain wall localized at the center $m=121$ of the chain ($N=241$ sites), as a function of the lattice site $r_n$ (a) of the ratio between the superimposed uniform field $b_0$ and the helical field $b_{yz}$, and (b) of the chemical potential (gate field).
The zero-energy peaks signal the presence of MBS localized alternatively at the edges, or at one edge and at the domain wall.
We assume a circular helimagnetic field with spatial frequency $\theta=2\pi/3$ superimposed with a uniform field $\mathbf{b}_0\parallel\hat{\mathbf{z}}$, and with $\Delta=t/2$, $b_{yz}=1.5t$, $\Delta\varphi_m=\pi/3$, $\mu=3t$ [panel (a)], and $b_0/b_{yz}=0.75$ (${{\delta b}^2/\meanbsq} \approx1$) [panel (b)].
}
	\label{fig:LDOS}
\end{figure}

To illustrate the formation of topological domains and the consequent pinning of MBS, we calculate the local density of states (LDOS) at zero energy, which corresponds to the probability to find an intragap MBS at a given position $r_n$ along the chain.
\Cref{fig:LDOS}(a) shows the LDOS in the case of a circular helimagnetic field with spatial frequency $\theta=2\pi/3$ superimposed with a uniform field, as a function of the ratio between the superimposed uniform field $b_0$ and the helical field $b$, with a fixed phase-difference $\Delta\varphi_m=\pi/3$ at the domain wall.
At $b_0=0$, the field amplitude is constant ($\delta b=0$) and thus the left and right domains are topologically equivalent [$\mathcal{P}(\varphi_L)=\mathcal{P}(\varphi_R)=-1$].
In this case MBS are necessarily localized at the two edges of the chain.
Nevertheless, as the magnitude of the applied field $b_0$ increases such that $\delta b>{\delta b}_c$, one of the MBS becomes pinned to the domain wall, leaving one side of the chain topologically trivial [$\mathcal{P}(\varphi_L)\neq\mathcal{P}(\varphi_R)$].

Moreover, the formation of contiguous topological domains can be obtained in proximized nanowires also by tuning the chemical potential via an applied gate field.
This is similar to the case of homogeneous topological superconductors with uniform magnetic fields~\cite{Lutchyn2010,Oreg2010}, where a change in the chemical potential (or magnetic field intensity) can turn the topological state from trivial to nontrivial.
For example, \cref{fig:LDOS}(b) shows the localization of MBS as a function of the chemical potential $\mu$.
MBS are localized alternatively at the left or right edge and at the domain wall for different values of $\mu$.

Hence, by controlling the external magnetic field or gate voltage, MBS can be moved from the edge of the chain to the domain wall.
The position of MBS can be revealed by a measure of the local differential conductance at zero bias, which is proportional to the LDOS at zero-energy, obtained via scanning tunneling microscopy (STM)~\cite{Mourik2012,Nadj-Perge2014}.
The pinning of MBS is topologically robust against magnetic and nonmagnetic disorder, as long as the perturbation is small, as verified in the \citesupp.

\paragraph*{Experimental implementation}

The implementation of the model proposed here requires the realization of amplitude-modulated magnetic textures with a periodicity comparable with the lattice parameter of the superconducting chain, in the presence of domain walls on a similar length scale.

A possible implementation is to realize a heterostructure where a proximized nanowire with strong SOC is contiguous to an antiferromagnetic material.
The magnetic field of an antiferromagnet is indeed dominated by a nonzero quadrupolar contribution~\cite{Guy1979,Astrov1996}, which results in a periodic field with the same periodicity of the ordering vector.
The mismatch between the ordering vector of the antiferromagnet and the lattice parameter of the nanowire will then result in an amplitude-modulated Zeeman field at the wire lattice sites.
Domain walls in antiferromagnets have a typical thickness of hundreds of atoms~\cite{Weber2003}:
The ensuing pinned MBS will be thus localized on a similar length scale.
Notice also that amplitude-modulated magnetic orders occur, e.g., in multiferroics~\cite{Kobayashi2009,Pregelj2012} and gadolinium compounds~\cite{Rotter2001,Good2005}, and in general in materials with competing ferromagnetic and antiferromagnetic exchanges and in the presence of anisotropic distortions which favors magnetization along a specific axis~\cite{Mochizuki2009}.
In these systems, where helical orders are generally coupled with a spontaneous electric polarization, electric fields or currents may be employed to induce a sliding of the helical spin texture~\cite{Yamasaki2008}. 
Other means may involve using spin-torque mechanisms or spin-currents to twist the phase-offset of the helical order ~\cite{Sonin2010,Zutic2004}.

Another possible implementation can be realized at larger length scales via artificial one-dimensional superlattices realized using nanoscale lithography design on ultrathin films~\cite{Tadjine2016} or quantum dot solids~\cite{Polini2013}, in the presence of nanomagnets.
This would allow a fine-tuning of the magnetic texture and domain wall properties by controlling the position and orientation of the nanomagnets.

In principle, an amplitude-modulated texture can be also achieved in topological Yu-Shiba-Rusinov chains~\cite{Nadj-Perge2013,Nadj-Perge2014} via an external magnetic field~\cite{Li2016}.
Notice that a realistic theoretical description of such system should take into account the effect of the external field on the magnetic texture, which would likely relax into a non-planar magnetic configuration, and the dependence of the ordering vector on the Fermi momentum~\cite{Braunecker2009PRB,Meng2013,Klinovaja2013,Vazifeh2013,Schecter2015}.
Moreover, a more fundamental challenge is the creation of stable domain walls in these systems via, e.g., magnetic impurities.

\paragraph*{Conclusions}

We have found that amplitude-modulated magnetic fields can induce the emergence of a topologically inhomogeneous state via the formation of topological domains.
In these systems, MBS become pinned to domain walls between contiguous and topologically-inequivalent domains.
Such systems can be in principle realized with proximized nanowires in a magnetic field induced by an antiferromagnet, in artificial one-dimensional superlattices via nanoscale lithography, or in quantum dot solids.

Remarkably, the formation of topological domains and the localization of MBS can be externally controlled by tuning an applied and uniform magnetic or gate field, without moving the domain wall.
Controlling the localization of MBS is the first step in the realization of a reliable braiding scheme in topological devices.

\paragraph*{Acknowledgments}
The authors thank Wei Chen, Paola Gentile, and Andreas Schnyder for useful discussions.
PM thanks Peter Stano, Daniel Loss, Jelena Klinovaja, and Alex Matos-Abiague for their valuable feedback and suggestions.



\clearpage\pagebreak\onecolumngrid\clearpage\thispagestyle{empty}

\begin{center}
\textbf{\large
Controlling Majorana states in topologically inhomogeneous superconductors:
\\[.5mm]
Supplemental Material
}

\quad

Pasquale Marra\hyperlink{mail}{$^*$} and Mario Cuoco
\\[.5mm]
{\it\small
CNR-SPIN, I-84084 Fisciano (Salerno), Italy
and
\\[-.5mm]
Department of Physics ``E.\,R.\, Caianiello'', University of Salerno, I-84084 Fisciano (Salerno), Italy
}

\end{center}

\setcounter{equation}{0}\setcounter{figure}{0}\setcounter{table}{0}\setcounter{section}{0}
\setcounter{page}{1}
\makeatletter
\renewcommand{\thepage}{S\arabic{page}}
\renewcommand{\theequation}{S\arabic{equation}}\renewcommand{\thefigure}{S\arabic{figure}}\renewcommand{\thetable}{S\arabic{table}}\renewcommand{\thesection}{S\arabic{section}}\renewcommand{\bibnumfmt}[1]{[S#1]}\renewcommand{\citenumfont}[1]{S#1}

\begin{center}\parbox{14cm}{\setlength{\parindent}{1em}
\indent\small
In this Supplemental Material we will establish under which conditions the energy spectrum and the topological invariant of the system described by Hamiltonian~\eqref{eq:Hamiltonian} of the main text can explicitly depend on the boundary phase-offset $\varphi$ of an amplitude-modulated Zeeman field $\mathbf{b}_n$.
}
\end{center}

\quad\vspace{3mm}

\twocolumngrid

\subsection{Unitary rotation of the spin basis}

Following Ref.~\onlinecite{Choy2011SM}, the Hamiltonian~\eqref{eq:Hamiltonian} of the main text can be transformed by employing a unitary rotation of the spin basis, which reads
\begin{equation}
U_n=
\begin{bmatrix}
\cos{(\theta_n/2)} & -\sin{(\theta_n/2)} e^{-i\phi_n} \\
\sin{(\theta_n/2)} e^{i\phi_n} & \cos{(\theta_n/2)}
\end{bmatrix}.
\label{eq:Rotation}
\end{equation}
where $\theta_n$ and $\phi_n$ are the inclination and azimuth of the magnetic field at the lattice site $r_n$, which can be written in general as
\begin{equation}
\mathbf{b}_n=b_n[\sin\theta_n \cos\phi_n, \sin\theta_n \sin\phi_n,\cos\theta_n].
\label{eq:Field}
\end{equation}
If the SOC vanishes ($\Omega=\mathbbm{1}$), the transformed BdG Hamiltonian reads
\begin{gather}
\mathcal{H}_\text{SO}=
\frac12\sum_n
\boldsymbol\Psi_n^\dag
\cdot\!
\begin{bmatrix}
2t-\mu+b_n{\sigma_z} & \imath\sigma_y\Delta\\
[\imath\sigma_y\Delta]^\dagger & -(2t-\mu+b_n{\sigma_z})
\end{bmatrix}
\!\cdot\!
\boldsymbol\Psi_n
\nonumber\\
-
\frac12\sum_n
\boldsymbol\Psi_n^\dag
\!\cdot\!
\begin{bmatrix}
t\Omega_n&0\\
0&-t\Omega_n
\end{bmatrix}
\!\cdot\!
\boldsymbol\Psi_{n+1}
+\text{H.c.},
\label{eq:HamiltonianRotated}
\end{gather}
where the unitary matrix $\Omega_n$ is given by
\begin{equation}
\Omega_{n}=U_{n}^{\dagger}U_{n+1}=
\begin{bmatrix}
\alpha_n & -\beta_n^* \\
\beta_n & \alpha_n^*
\end{bmatrix},
\end{equation}
with
\begin{align}
\alpha_n &= \cos{\frac{\theta_n}{2}}\cos{\frac{\theta_{n+1}}{2}} + \sin{\frac{\theta_n}{2}}\sin{\frac{\theta_{n+1}}{2}} e^{-\imath (\phi_n - \phi_{n+1})},
\nonumber\\
\beta_n &= \cos{\frac{\theta_n}{2}}\sin{\frac{\theta_{n+1}}{2}} e^{\imath\phi_{n+1}}-\sin{\frac{\theta_n}{2}}\cos{\frac{\theta_{n+1}}{2}} e^{\imath\phi_n}.
\end{align}
In the transformed Hamiltonian the Zeeman term becomes diagonal in the spin basis (the effective field $b_n\hat{\mathbf{z}}$ is along the $z$ axis), the hopping term is renormalized ($\propto t\alpha_n$), whereas an effective SOC appears ($\propto t\beta_n$).
It is straightforward to see that this effective SOC term $\propto t|\beta_n|$ is nonzero only if the field direction is not constant (noncollinear field).
If the field direction is constant and the intrinsic SOC vanishes, one has instead $\Omega_n=U_{n}^{\dagger}U_{n+1}=U_{n}^{\dagger}U_{n}=\mathbbm{1}$, i.e, $\alpha_n=1$ and $\beta_n=0$.

\subsection{Field with constant amplitude}

A global rotation of the spin basis is described by the unitary transformation
\begin{equation}
U(\Theta,\Phi)=
\begin{bmatrix}
\cos (\Theta/2) & -\sin(\Theta/2) e^{-\imath\Phi} \\
\sin (\Theta/2) e^{\imath\Phi} & \cos(\Theta/2)
\end{bmatrix},
\label{eq:GlobalRotation}
\end{equation}
which is equivalent to the introduction of the phase-offsets $\theta_n\to\theta_n+\Theta$ and $\phi_n\to\phi_n+\Phi$ in the field in \cref{eq:Field}.
Therefore any phase-offset which affects only the modulation of the field direction can be absorbed by the unitary transformation in \cref{eq:GlobalRotation} for some values of the angles $\Theta$ and $\Phi$.
For example, one can consider the case of a noncollinear Zeeman field with constant amplitude and with direction rotating in the $yz$ plane, i.e., with $\theta_n=n\theta+\varphi$ and $\phi_n=\pi/2$ in \cref{eq:Field} which gives
\begin{equation}
\mathbf{b}_n=b [\sin{(n\theta+\varphi)}\hat{\mathbf{y}}+\cos{(n\theta+\varphi)}\hat{\mathbf{z}}].
\end{equation}
In this case the phase-offset $\varphi$ is unitarily absorbed by a global rotation $U(\Theta,\Phi)$ with $\Theta=\varphi$ and $\Phi=\pi/2$.
Moreover, the effective Zeeman field in Hamiltonian~\eqref{eq:HamiltonianRotated} is uniform, being $b_n\hat{\mathbf{z}}=b\hat{\mathbf{z}}$, while the hopping term and SOC become respectively $\alpha_n=\cos\theta$ and $\beta_n =\imath\sin{\theta}$, and therefore one has
\begin{equation}
\Omega_{n}=\cos{(\theta/2)}+\imath\sigma_x\sin{(\theta/2)}
\end{equation}
The effective Zeeman field and the effective SOC do not depend on the phase-offset $\varphi$.
As a consequence, the energy spectrum and any other physical properties are not affected by a change of the phase-offset $\varphi$.
The phase-offset is immaterial in this case, since it is absorbed by a unitary transformation (rotation of the spin basis).

\subsection{Elliptic helimagnetic field}

An amplitude-modulated field can be realized by an elliptical helimagnetic field in the form
\begin{equation}
\mathbf{b}_n=b_y\sin{(n\theta+\varphi)}\hat{\mathbf{y}}+b_z\cos{(n\theta+\varphi)}\hat{\mathbf{z}},
\label{eq:FieldNoncircular}
\end{equation}
where $b_y\neq b_z$.
The Zeeman field is periodic in space with period $L=2\pi/\theta$, and we assume that the periodicity is commensurate to the lattice, i.e., that $L$ is an integer number.
One can immediately verify that the modulus square of the Zeeman field is given in this case by
\begin{equation}
b_n^2=\frac{b_z^2+b_y^2}2 + \frac{b_z^2-b_y^2}2 \cos{[2(n\theta+\varphi)]},
\end{equation}
which is not constant along the chain if $b_y\neq b_z$, but periodically modulated in space with period $L/2$ (the period is halved in this particular case).
The amplitude-modulation does explicitly depends on the phase-offset $\varphi$ and it is periodic in the phase-offset with period $\Delta\varphi=\theta/2$.
Notice that the average field and its average amplitude over the whole interval $n=0,L-1$ do not depend on the phase-offset, and are given by
\begin{equation}
\langle \mathbf{b}_n\rangle\!=\!\frac1L\!\sum\limits_{n=0}^{L-1}\! \mathbf{b}_n\!=\!0,
\quad
\langle {b}_n^2\rangle \!=\!\frac1L\!\sum\limits_{n=0}^{L-1}\! b_n^2 \!=\!\frac{b_y^2+b_z^2}2,
\end{equation}
as one can show by using the Euler identities and the geometric sum formula which gives in this case
\begin{equation}
\sum\limits_{n=0}^{L-1} e^{\imath(n\theta+\varphi)}= e^{\imath \varphi} \frac{1-e^{\imath L\theta}}{1-e^{\imath \theta}}=0.
\label{eq:Geometric}
\end{equation}
Therefore the amplitude of the Zeeman field can be written as
\begin{align}
b_n=\sqrt{\meanbsq+{\delta b}^2\cos{[2(n\theta+\varphi)]}},
\end{align}
where $\meanbsq=\langle {b}_n^2\rangle$ and ${\delta b}^2=(b_{z}^2-b_y^2)/2$.

\begin{figure}[t]
\centering
\includegraphics[scale=1,resolution=600]{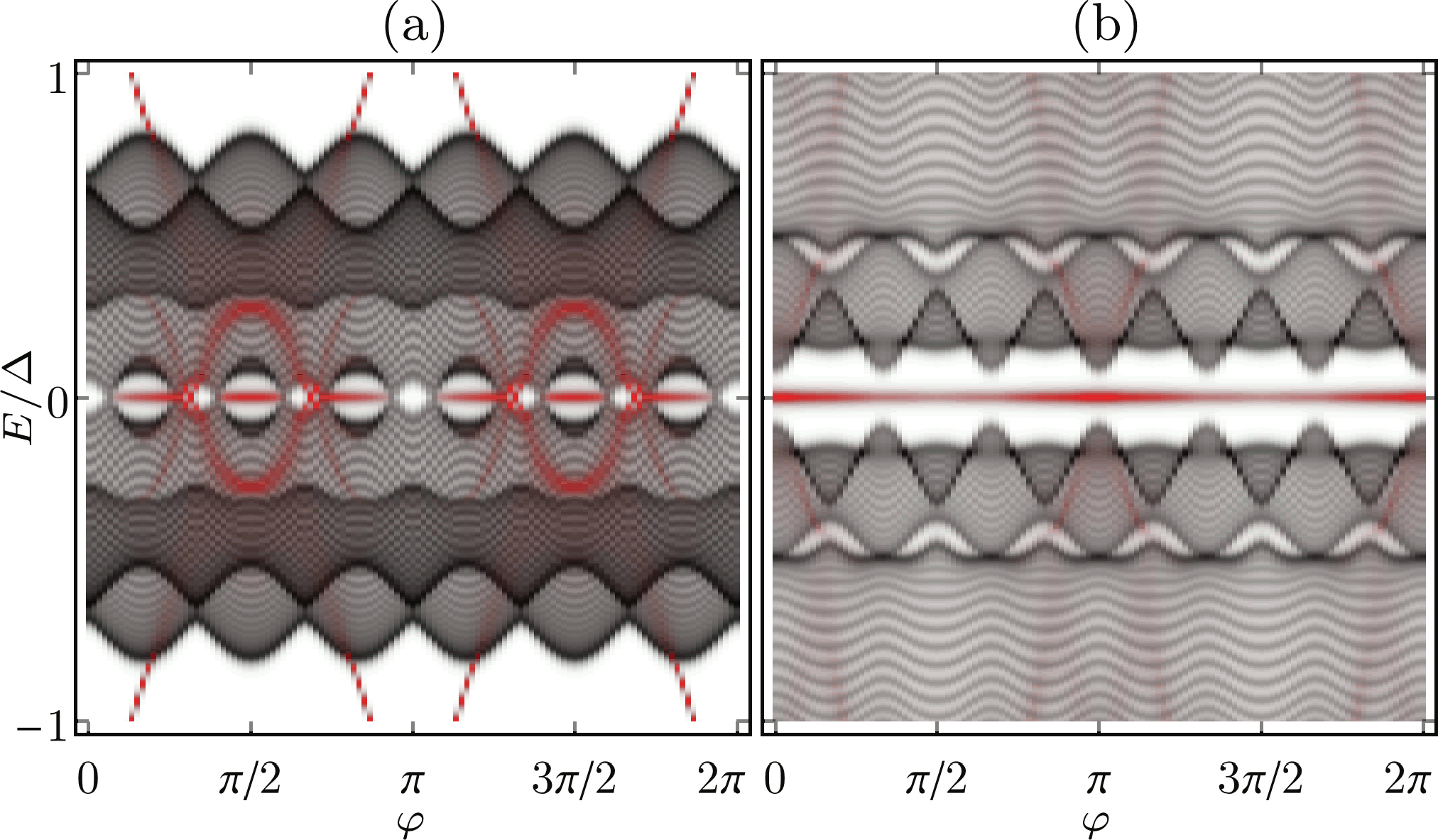}\\[-1mm]
	\caption{%
DOS in the bulk (dark) and at the edges (color) in the low-energy range as a function of the boundary phase-offset $\varphi$ calculated in the case of an elliptical helical field with spatial frequency $\theta=2\pi/3$ [as in \cref{fig:System}(a) of the main text].
(a) Nontrivial gaps with MBS alternating with trivial gaps as a function of the phase-offset $\varphi$ obtained for $\mu=4t$.
(b) Nontrivial gap with MBS in the whole range $\varphi\in[0,2\pi]$ obtained for $\mu=2.5t$.
Notice that in both cases the bulk energy levels are periodic in the phase-offset with period $\Delta\varphi=\theta/2=\pi/3$.
We assume $\Delta=t/2$, $b_y=2t$, and $b_z=0.4t$, which give ${\delta b}^2/\meanbsq \approx0.9$.
}
	\label{fig:Edges-Elliptic}
\end{figure}

\Cref{fig:Edges-Elliptic} shows the energy spectra in the presence of an elliptical helimagnetic field with $\theta=2\pi/3$.

\subsection{Circular helimagnetic field with superimposed uniform field}

An amplitude-modulated field can be also realized by a circular helimagnetic field superimposed with a coplanar uniform field, in the form
\begin{equation}
\mathbf{b}_n=b_{yz} \left[\sin{(n\theta+\varphi)}\hat{\mathbf{y}}+\cos{(n\theta+\varphi)}\hat{\mathbf{z}}\right]+\mathbf{b}_0,
\label{eq:FieldCircularPerturbed}
\end{equation}
where $\mathbf{b}_0=b_{0y}\hat{\mathbf{y}}+b_{0z}\hat{\mathbf{z}}$ lies in the $yz$ plane.
The Zeeman field is periodic in space with period $L=2\pi/\theta$, and we assume again that the periodicity is commensurate to the lattice, i.e., that $L$ is integer.
In this case the modulus square of the Zeeman field is given by
\begin{align}
b_n^2=b_{yz}^2 + b_0^2 + 2b_{yz} b_0\cos{(n\theta+\varphi)},
\end{align}
if one redefines $\varphi\to\varphi+\theta_{yz}$ where $\tan\theta_{yz}=b_{0y}/b_{0z}$.
Therefore the field amplitude is not constant along the chain if $b_0\neq 0$, but periodically modulated in space with period $L$.
Also in this case the amplitude-modulation does explicitly depends on the phase-offset $\varphi$, and it is periodic in the phase-offset with period $\Delta\varphi=\theta$.
Again, the average field and its average amplitude over the whole spatial period $n=0,L-1$ do not depend on the phase-offset, and are given in this case by
\begin{equation}
\langle \mathbf{b}_n\rangle\!=\!\frac1L\!\sum\limits_{n=0}^{L-1}\! \mathbf{b}_n\!=\!\mathbf{b}_0,
\quad
\langle \mathbf{b}_n^2\rangle\!=\!\frac1L\!\sum\limits_{n=0}^{L-1}\! b_n^2 \!=\!b_{yz}^2+b_0^2,
\end{equation}
as one can show by using the Euler identities and the geometric sum formula in \cref{eq:Geometric}.
Therefore the amplitude of the Zeeman field can be written as
\begin{align}
b_n=\sqrt{\meanbsq+\delta b^2\cos{(n\theta+\varphi)}},
\end{align}
where $\meanbsq=\langle {b}_n^2\rangle$ and $\delta b^2=2 b_{yz} b_0$.

\subsection{Amplitude-modulated field: General case}

In the two cases considered in \cref{eq:FieldNoncircular,eq:FieldCircularPerturbed}, the field amplitude is not constant, but periodically modulated along the chain with a period which is a submultiple of $\Delta\varphi=2\pi/\theta$.
Moreover, the phase-offset $\varphi$ cannot be absorbed by the unitary rotation $U_n$ of the spin basis, since the effective field amplitude $b_n$ of the transformed Hamiltonian~\eqref{eq:HamiltonianRotated} does depend explicitly on the phase-offset.
Indeed, an amplitude-modulated field can be written in general as
\begin{equation}
\mathbf{b}_n=f(n\theta+\varphi)[\sin\theta_n \cos\phi_n, \sin\theta_n \sin\phi_n, \cos\theta_n],
\label{eq:AMField}
\end{equation}
where $f(x)$ is a periodic and nonconstant function of period $2\pi$ (or submultiples).
\Cref{eq:FieldNoncircular,eq:FieldCircularPerturbed} are special cases of \cref{eq:AMField}.
In all cases where the Zeeman field can be written in the form of \cref{eq:AMField}, the field amplitude is modulated along the chain and the phase-offset $\varphi$ cannot be absorbed by unitary rotations of the spin basis.
Nevertheless, in a continuous system, the phase-offset $\varphi$ can be always absorbed by a translation $n\to n-\varphi/\theta$ in \cref{eq:AMField} [as well as in \cref{eq:FieldNoncircular,eq:FieldCircularPerturbed}].
In a discrete system instead, this is not possible, unless such translation coincide with a discrete translation of the lattice by a multiple of the lattice constant, i.e., if $\varphi/\theta$ is an integer.
For the same reason one can see that the Zeeman field in \cref{eq:AMField} [or in \cref{eq:FieldNoncircular,eq:FieldCircularPerturbed}] is periodic in the phase-offset $\varphi$ with period $\Delta\varphi=\theta$ (see also Ref.~\onlinecite{Marra2015SM}) up to discrete translations of the lattice, and therefore one can always assume that $\varphi\in[0,\theta]$.
Consequently, in the presence of amplitude-modulated Zeeman fields, the energy spectrum and other physical properties, e.g., the topological invariant, do depend on the phase-offset $\varphi$.

\subsection{Topological invariant}

The $\mathbb{Z}_2$ topological invariant can be defined as $\mathcal{P}=\sgn\left[\prod_{k=-k}F(k)\right]$, where the product is extended over the time-reversal invariant momenta $k=-k$ and where $F(k)=\pf\left[H(k)\imath\tau_x\right]$ is the Pfaffian of the antisymmetric matrix $H(k)\imath\tau_x$, being $H(k)$ the BdG Hamiltonian in momentum space and $\tau_x$ the first Pauli matrix in particle-hole space.
Here we will determine the dependency of the topological invariant on the phase-offset $\varphi$ of the field modulation.
Since, as we have seen, a system with noncollinear Zeeman field is unitarily equivalent to a system with collinear field and effective SOC, we will calculate the topological invariant only in the latter case.
Hence, let us consider a system with a collinear amplitude-modulated magnetic field with $\mathbf{b}_n=[\meanb+\delta b \cos{(n\theta+\varphi)}]\hat{\mathbf{z}}$ with finite and uniform SOC $\lambda\neq0$.
In this case, Hamiltonian~\eqref{eq:Hamiltonian} of the main text can be rewritten in momentum space as
\begin{equation}
\widetilde{\cal H}
=\frac12\sum_{k}
\boldsymbol\Psi_k^\dag \cdot h(k) \cdot \boldsymbol\Psi_k^\nodag
+
\boldsymbol\Psi_k^\dag \cdot w \cdot \boldsymbol\Psi_{k+\theta}^\nodag
+\text{H.c.},
\label{eq:Hamiltoniank}
\end{equation}
where the matrices $h(k)$ and $w$ represent respectively the unperturbed Bloch Hamiltonian in the case of $\delta b=0$ and the coupling at different momenta $k'-k=m\theta$ with $m\in\mathbb{Z}$ ensuing from the presence of a finite amplitude-modulation of the field.
They are defined as
\begin{widetext}
\begin{equation}
h(k)=
\begin{bmatrix}
(2t-\mu-2t\cos{k})+2\lambda t\sin{k}\,\sigma_y + \meanb\sigma_z &
\hspace{-10mm}
\Delta\imath\sigma_y\\
\hspace{-20mm}
(\Delta\imath\sigma_y)^\dag&
\hspace{-28mm}
-(2t-\mu-2t\cos{k})
-2\lambda t\sin{k}\,\sigma_y
-\meanb\sigma_z
\end{bmatrix},
\qquad
w=
\dfrac{e^{\imath\varphi}}2
\begin{bmatrix}
\delta b\sigma_z&0\\
0&-\delta b \sigma_z
\end{bmatrix}.
\end{equation}
Moreover, if the periodicity of the field is commensurate with the lattice, Hamiltonian~\eqref{eq:Hamiltoniank} can be written in matrix form as
\begin{equation}
\widetilde{\cal H}
=\frac12\sum_{k}
\left[\boldsymbol\Psi_k^\dag, \boldsymbol\Psi_{k+\theta}^\dag, \ldots, \boldsymbol\Psi_{k+(q-1)\theta}^\dag\right]
\cdot
\begin{bmatrix}
h(k)&w\\
w^\dagger&h(k+\theta)\\
&&\ddots\\
&&&h(k+(q-1)\theta)\\
\end{bmatrix}
\cdot
\begin{bmatrix}
\boldsymbol\Psi_k^\nodag\\ \boldsymbol\Psi_{k+\theta}^\nodag\\ \vdots \\\boldsymbol\Psi_{k+(q-1)\theta}^\nodag
\end{bmatrix},
\label{eq:HamiltoniankMatrix}
\end{equation}
\end{widetext}
with $\theta = 2\pi p/q$ with $p,q\in \mathbb{Z}$ coprimes.
The Hamiltonian is periodic in the momentum with period $\theta$.
Indeed, a direct substitution $k\to k+\theta$ in Hamiltonian~\eqref{eq:HamiltoniankMatrix} coincide with a mere rearrangement of the order of the matrix blocks.
The periodicity in the momentum results in a reduced Brillouin zone given by $[0,2\pi/q]$.
Notice that in this case the two time-reversal invariant points are respectively $k=0$ and $k=\pi/q$, since the momentum $k=\pi/q$ is invariant under time-reversal ($k\to-k$) due to the periodicity of the reduced Brillouin zone, i.e., $-k=-\pi/q=\pi/q-2\pi/q\equiv\pi/q$.

If the amplitude-modulation vanishes ($\delta b=0$), one has $w=0$ and therefore Hamiltonian~\eqref{eq:HamiltoniankMatrix} is a diagonal block matrix of $q$ blocks $h(k+m\theta)$ with $m=0,q-1$.
To calculate $F(k)$, we notice that the absolute value of the Pfaffian of an antisymmetric matrix is equal to the square root of its determinant, and that the determinant of a block matrix is given by the product of the determinants of each block.
Therefore one obtains
\begin{equation}
F(k)=F_0(k)\equiv
\pm\sqrt{\prod\limits_{m=0}^{q-1} \det{\left[h(k+m\theta)\right]}},
\label{eq:PfaffianExpansion0}
\end{equation}
for $k=0,\pi/q$, where
\begin{gather}
\det{\left[h(k+m\theta)\right]}=
\left[
\Delta^2+\mu^2-b^2
-4 \mu t-2t^2\left(\lambda^2-3\right)
\right.\nonumber\\\left.
+4t(\mu-2t)\cos{k}
+2t^2\left(\lambda^2+1\right) \cos{(2k)}
\right]^2.
\end{gather}
Instead, if the field is amplitude-modulated ($\delta b\neq0$), one has $w\neq0$ and therefore the Pfaffian expansion in \cref{eq:PfaffianExpansion0} acquires additional terms which depends on the phase-offset $\varphi$.
We recall that the physical properties of the system are periodic in the phase-offset with period $\theta=2\pi p/q$.
Therefore the additional terms in the Pfaffian expansion can be written as a Fourier sum of a finite number of harmonics with periods $m q$ and proportional to $\delta b^{m q}$ where $m\in\mathbb{Z}$.
Neglecting higher harmonics, one obtains
\begin{equation}
F(k)\approx F_0(k) + C{\delta b}^{q} \cos{(q\varphi+\varphi')},
\label{eq:PfaffianExpansion1}
\end{equation}
for $k=0,\pi/q$, where $C$ is a global prefactor and a phase-shift which depends on the magnetic field, chemical potential, superconducting pairing, and periodicity of the field.
Since a system with collinear amplitude-modulated field with uniform SOC is symmetric under spatial inversion, one can conclude that the topological invariant must be symmetric under the inversion of the phase-offset $\varphi\to-\varphi$ and therefore the phase-shift $\varphi'$ in \cref{eq:PfaffianExpansion1} can be neglected being $\varphi'=0$ or $\pi$.
Hence one obtains \cref{eq:Pfaffian} of the main text.

\subsection{Robustness}

\begin{figure}
\centering
\includegraphics[scale=1,resolution=600]{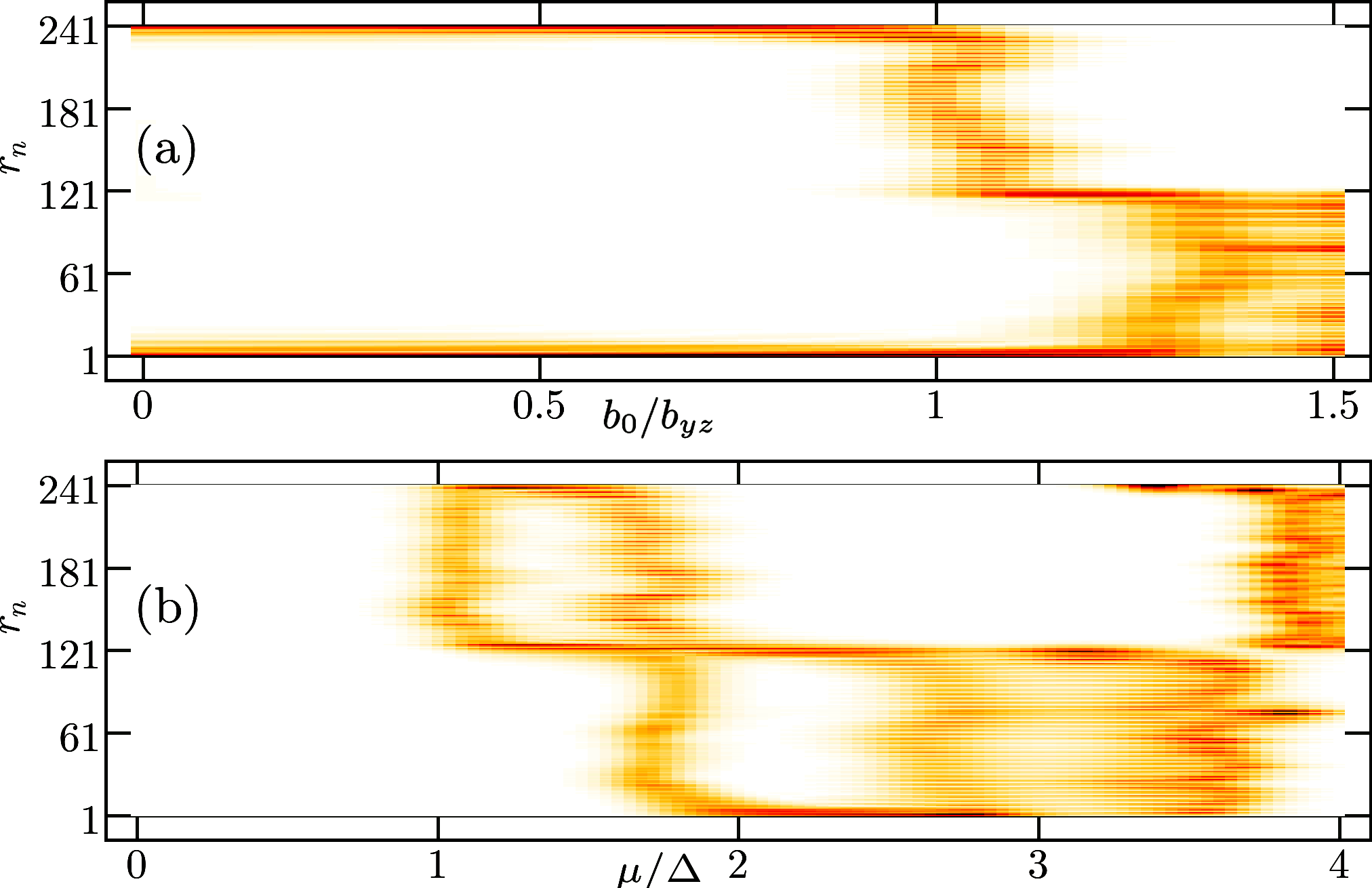}\\[-1mm]
	\caption{%
LDOS at zero energy in the presence of a domain wall localized at the center of the chain, as a function of the lattice site $r_n$ and (a) of the ratio between the superimposed uniform field $b_0$ and the helical field $b_{yz}$, and (b) of the chemical potential (gate field) as in \cref{fig:LDOS} of the main text, but in the presence of disorder.
The pinning of MBS to the domain wall is topologically robust against disorder, as long as the strength of the perturbation is small.
In particular we assume that $b^\text{imp}/b_{yz}=0.1$ and $\varepsilon^\text{imp}/t=0.1$ [see Hamiltonian~\eqref{eq:HamiltonianDisorder}].
}
	\label{fig:LDOSDisorder}
\end{figure}

We consider here the case where the Hamiltonian~\eqref{eq:Hamiltonian} of the main text is perturbed by disorder.
We introduce disorder as an additional Hamiltonian term $\mathcal{H}_\text{imp}$ which describes both magnetic and nonmagnetic random perturbations, and which in particle-hole space reads
\begin{equation}
\mathcal{H}_\text{imp}=
\frac12\sum_n
\boldsymbol\Psi_n^\dag
\!\cdot\!
\begin{bmatrix}
\mathbf{b}_n^\text{imp}\!\cdot\!\boldsymbol{\sigma}-\varepsilon_n^\text{imp}&0\\
\hspace{-1cm}
0
&
\hspace{-1.2cm}
-\mathbf{b}_n^\text{imp}\!\cdot\!\boldsymbol{\sigma}^*+\varepsilon_n^\text{imp}
\end{bmatrix}
\cdot
\boldsymbol\Psi_n,
\label{eq:HamiltonianDisorder}
\end{equation}
where $\mathbf{b}_n^\text{imp}$ and $\varepsilon_n^\text{imp}$ are taken to be randomly distributed uniformly with maximum amplitudes $b^\text{imp}\ll \meanb$ and $\varepsilon^\text{imp}\ll t$.

\Cref{fig:LDOSDisorder} shows the effect of the disorder to the LDOS at zero energy in the case of a circular helimagnetic field with spatial frequency $\theta=2\pi/3$ superimposed with a uniform field $\mathbf{b}_0$ and with a domain wall localized at the center of the chain (cf.\ \cref{fig:LDOS} of the main text).
In particular we assume that $b^\text{imp}/b_{yz}=0.1$ and $\varepsilon^\text{imp}/t=0.1$.
As one can see, the pinning of MBS to the domain wall is not influenced by the presence of disorder.


\hypertarget{mail}{}

\end{document}